\documentclass[a4paper,10pt]{article}  

\usepackage{graphicx}

\usepackage{ulem}
\usepackage{cancel}
\usepackage{amsmath}
\usepackage{amssymb}
\usepackage{color}
\usepackage[usenames,dvipsnames]{xcolor}
 
\usepackage[abbr]{harvard}
\citationstyle{dcu}
\citationmode{abbr}
\renewcommand{\harvardand}{\&}

\newcommand{\EXP}[1]{\mathrm{e}^{#1}} 
\newcommand{\DEF}{\stackrel{\mathrm{def}}{=}}

\newcommand{\imat}{{\mathrm{i}}} 
\newcommand{\dmat}{\mathrm{d}}

\newcommand{\omat}{\mathrm{o}}
\newcommand{\goe}{\mathfrak{e}}

\newcommand{\gom}{\mathfrak{m}}

\newcommand{\gru}{\bold{u}}
\newcommand{\grv}{\bold{v}}
\newcommand{\grw}{\bold{w}}
\newcommand{\grT}{\bold{T}}
\newcommand{\im}{\operatorname{Im}}
\newcommand{\re}{\operatorname{Re}}

\newcommand{\fatops}[2]{\genfrac{}{}{0pt}{2}{#1}{#2}}   
\newcommand{\scl}{\fatops{ \raisebox{-.3cm}{$\textstyle\sim$} }
                         { \scriptstyle\hbar\to 0             }
                 }

\newcommand{\ket}[1]{|#1\rangle} 
\newcommand{\kete}[1]{|\kern.3ex#1\kern.3ex\rangle}

\newcommand{\brae}[1]{\langle\kern.3ex #1 \kern.3ex|}
\newcommand{\braket}[2]{\langle #1 |#2\rangle}
\begin{document}

\title{A primer for resonant tunnelling}
\author{J\'er\'emy Le Deunff$^2$, Olivier Brodier$^1$ \& Amaury Mouchet$^1$\\ \\
$^1$Laboratoire de Math\'ematiques 
  et Physique Th\'eorique,\\ Universit\'e Fran\c{c}ois Rabelais de Tours 
--- \textsc{\textsc{cnrs (umr 6083)}},\\
F\'ed\'eration Denis Poisson,\\
 Parc de Grandmont 37200
  Tours,  France. \\{brodier, mouchet @lmpt.univ-tours.fr} \\ \\
   {$^2$}Max-Planck-Institut f{\"u}r Physik komplexer Systeme, \\
   N{\"o}thnitzer Stra{\ss}e 38, \\
   01187 Dresden, Germany. \\{deunff@pks.mpg.de}}
%\author{Amaury Mouchet}
%\address{Laboratoire de Math\'ematiques 
% et Physique Th\'eorique,\\ Universit\'e Fran\c{c}ois Rabelais de Tours 
%--- \textsc{\textsc{cnrs (umr 6083)}},\\
%F\'ed\'eration Denis Poisson,\\
%Parc de Grandmont 37200
%  Tours,  France.}
%\ead{mouchet@phys.univ-tours.fr}

\date{\today}

\maketitle 


\begin{abstract} Resonant tunnelling is studied numerically and analytically with the help
of a three-well quantum one-dimensional time-independent model.  The simplest cases are considered
where the three-well potential is polynomial or piecewise constant.
\end{abstract}

\section{Introduction}

Even though the expression ``tunnel effect'' was coined for the first
time in 1931 by Schottky in German,
\citeaffixed{Merzbacher02a}{``Wellenmechanische Tunneleffekt'',
  according to}, the importance of the quantum transmission through a
potential barrier at an energy below its maximum was acknowledged
immediately after quantum mechanics reached its maturity in 1926.
Indeed, Hund published the first papers on tunnelling in 1927 where a
one-dimensional double-well was introduced to modelise chemical
binding \cite{Hund27a} and the deformation potential of a
$\mathrm{NH}_3$-like molecule \cite{Hund27b}. In the latter article, Hund obtained the crucial point
that tunnelling was exponentially sensitive to the characteristic parameters of the barrier. The very
same year, tunnelling was first considered in unbounded models by
\citeasnoun{Nordheim27a} to describe the electronic emission from
metals. One year later, \citeasnoun{Fowler/Nordheim28a} successfully
described how this emission was driven by high electric field and
\citeasnoun{Gamow28a}, \citeasnoun{Gurney/Condon28a} showed how
tunnelling was involved in the $\alpha$-decay of some radioactive
nuclei; in both cases, the exponential dependence of the decay rates
with energy was explained by tunnelling through a one-dimensional
barrier that modelled the mean attractive field created respectively
by the metal and the nucleons (and the latters were still unknown in
1928).

Such tunnelling models can be introduced very early in a first course
on quantum mechanics. As soon as the stationnary Schr\"odinger
equation is presented together with the constraints imposed on its
resolution (normalisation, smoothness of the solutions), tunnelling
can be understood as a direct consequence of the wavy nature of
quantum particles. For usual textbooks treatments see for instance
\cite[\S~11.5]{Bohm51a}, \cite[chap.~III, \S~7]{Messiah59a},
\cite[chaps.~5 and~6]{Merzbacher70a}. Probably more striking for the
beginners than the quantization of the energy levels of a bounded
system, tunnelling provides a first contact with the strangeness of
quantum phenomena, even for those already familiar with evanescent
waves in optics.

Tunnelling has nowadays been extended to qualify \textit{any quantum
  process that is forbidden by real solutions of classical
  equations}\footnote{For more than one degree of freedom or for
  time-dependent systems, the interdiction may come from constraints
  other than the energy conservation. If we look at the dynamics in
  the phase-space and make the canonical transformation that exchanges
  position with momenta, it also appears that the reflexion above a
  barrier can be considered as a tunnelling process.}; thus, since a
quantum/classical comparison is somehow necessary, it is, by
definition, a phenomenon that occurs in the semiclassical limit, where
formally~$\hbar\to0$ (physically this corresponds to large classical
actions compared to the Planck constant. This limit is not trivial : from the Schr\"odinger equation
 we can see that the value $\hbar=0$ is singular; however, in practice,
 the semiclassical approximation works quite well even 
for relatively large value of $\hbar$. The only requirement 
in our case is that $\hbar$ must remain small enough to get two 
almost degenerate eigenvalues below the tunnelling barrier.) With the help of approximate
methods of resolution of ordinary differential equation that were
originally developed by \citeasnoun{Jeffreys25a},
\citeasnoun{Kramers26a}, \citeasnoun{Brillouin26a},
\citeasnoun{Wentzel26a}---the so-called \textsc{jwkb} approximation
that provide $\hbar$ asymptotic expansions for the quantum quantities
\citeaffixed[chap.~10]{Bender/Orszag78a}{for a rigourous though
  accessible presentation with more details than the ones usually
  presented in quantum mechanics textbooks see}---one may also gain
some insight of how subtle the transition between the classical and
the quantum world occurs. However, from the singular behaviour of the
wave functions (in their semiclassical amplitudes near the classical
turning points and, in any case, in their phases), it appears
immediately that the semiclassical limit~$\hbar\to0$ is not as simple
as the other familiar limit $c\to\infty$ that describes the transition
towards non-relativistic theories (physically this corresponds to
small classical velocities compared to the speed of light~$c$).
\begin{figure}[!ht]
\center \includegraphics[width=\textwidth]{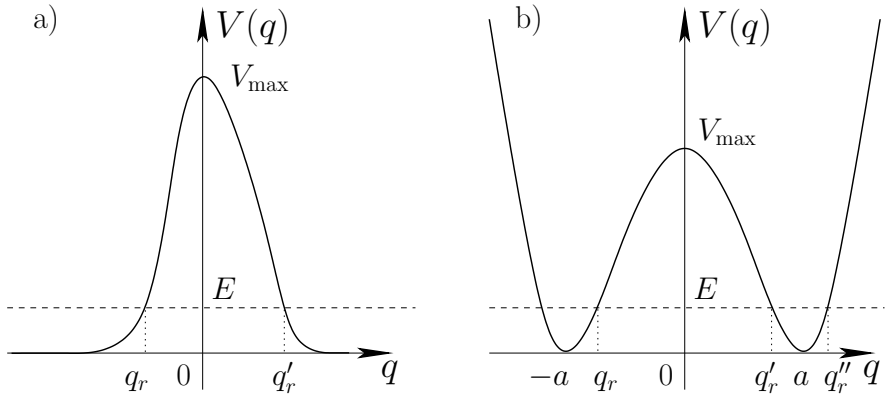}
\caption{\label{fig:barrieres}a) One simple potential barrier and
    b) the double-well potential provide the first illustrations of
    tunnelling in one dimension.}
\end{figure}

Remaining at an introductory level, within \textsc{jwkb}
approximation, it is quite simple to justify the exponential form
taken by the transmission factor $\tau$ (defined to be the square of
the ratio between the amplitudes of the outgoing and incoming waves) of one barrier like the one depicted
in figure~\ref{fig:barrieres}a) \cite[\S~12.13]{Bohm51a}, \cite[\S~50,
  eq.~(50.5)]{Landau/Lifshitz58c}, \cite[chap.~VI, \S~10 and
  exercise~5]{Messiah59a} \cite[chap.~7, eq.~(7.33)]{Merzbacher70a}
\begin{equation}\label{eq:tau}
\tau \scl \alpha\, \EXP{-2A/\hbar}\;,
\end{equation}
or the tunnelling beating period~$T=\hbar/\Delta E$ in a symmetric
double-well like the one given in figure~\ref{fig:barrieres}b),
\cite[\S~50, problem 3]{Landau/Lifshitz58c}. In that latter case,
tunnelling appears as a coupling between two states localised in the
two wells that lifts the degeneracy of their energy. Actually, below
the maximum of the barrier~$V_{\mathrm{max}}$ the symmetric and
antisymmetric exact eigenvalues come by pairs~$E_n^+$ and $E_n^-$
labelled by a finite number of integers~$n=0,1,...,N$ and the
splitting of the~$n$th doublet is given
by \begin{equation}\label{eq:deltaE} \Delta E_n\DEF|E_n^+-E_n^-| \scl
  \alpha\, \hbar\,\EXP{-A/\hbar}\;.
\end{equation}
It requires some skill to get the correct prefactors~$\alpha$
\cite{Garg00a} but if we do not bother about them, the above
expressions can be straightforwardly obtained. In both cases~$A$ is
the classical action ``under the barrier $V(q)$" of the particle with
mass~$m$, more precisely the imaginary part of the action computed
with one branch of the imaginary momentum: \begin{equation}
  A(E)\DEF\int^{q'_r(E)}_{q_r(E)}\sqrt{2m\big(V(q)-E\big)}\,\dmat q\;,
\end{equation}
where $q_r$ and $q'_r$ are the positions of the turning points at the
energy~$E$ at which the barrier penetration occurs. In the double well
case, $E$ is the average energy of the tunnelling doublet
$E_n=(E_n^++E_n^-)/2$ and may be determined semiclassically within one
well by the Einstein-Brillouin-Keller quantization condition
\cite[\S~11.5]{Bohm51a}, \cite[\S~48]{Landau/Lifshitz58c},
\cite[chap.~VI,\S~11]{Messiah59a}, \cite[chap.~7,\S~3]{Merzbacher70a},
\begin{eqnarray}
\oint_{\substack{\\[1em] \text{closed 
trajectory}\\ \text{ at energy $E$}}}\hspace{-1.5cm} p(q)\,\dmat q &=&2\pi\hbar\Big(n+\frac{1}{2}\Big)\;\label{eq:ebk}\\
 &=&
  2\int^{q''_r(E)}_{q'_r(E)}\sqrt{2m\big(E-V(q)\big)}\,\dmat q\;,
\end{eqnarray}
where $q''_r(E)$ stands for the most right turning point at energy~$E$.

Simple exponential laws like~$\eqref{eq:tau}$ or $\eqref{eq:deltaE}$,
discovered by the pioneers of tunnelling mentioned above, are
advantageously used in many applications, most notably the principle
of the scanning tunnelling microscope (or rather nanoscope)
\cite{Eigler/Schweizer90a}. However, when the potential has a richer
structure than just one bump, even when staying in one dimension,
strong deviations by several order of magnitude on the transmission or
on the splitting are to be expected. For instance, even if the
transmission of one barrier can be made arbitrarily small because
of~\eqref{eq:tau}, the total transmission factor of two such
successive bumps may be enhanced up to its maximum value~1. This
resonance tunnelling has fundamentally the same roots than the
resonance scattering by one well \cite[chap.~11,\S~7]{Bohm51a},
\cite[chap.~III, \S~6]{Messiah59a} \cite[chap.~VI,
  \S~8]{Merzbacher70a}: an enhancement is produced by the constructive
interference of waves reflected back and forth in between the two
bumps. In optics such a coherent superposition of ``trapped''
electromagnetic waves is the principle of the Fabry-P\'erot
interferometer \cite[\S~7.6]{Born/Wolf80a}. Though accessible with no
more technical nor conceptual tools than the ones involved in the
\textsc{jwkb} derivation of~$\eqref{eq:tau}$, $\eqref{eq:deltaE}$ and
\eqref{eq:ebk}, quantum analogous treatment of the Fabry-P\'erot
effect is scarcely treated in the introductory literature
\cite[chap.~12, \S\S~14-17]{Bohm51a}, see \possessivecite{Granot06a}
recent publication in this journal. The aim of the present article is
to fill a gap and complete the study of resonant tunnelling by
considering the spectrum of bounded systems rather than
scattering. Multidimensionnal tunnelling (including time-dependent
one-dimensional systems) requires theoretical tools far beyond the
readership's interest of the present journal and is still the subject
of an intense research field. However, recent studies has shown that
resonant tunnelling (mainly in bounded systems) paved the road to a
better understanding of tunnelling in complex systems
\cite{Schlagheck+11a}. For instance, the huge fluctuations by several
orders of magnitude in some tunnelling doublets are precisely of the
same origin than the resonance that will be carefully described in the
present article. We will remain with a one-dimensional static
potential~$V$ and choose two tractable models that involve three
wells. First
(section~\ref{sec:q6pot}) we will consider a smooth potential, the
simplest one being given by a polynomial of degree~6. One of its
advantage is that the spectrum can be numerically computed with very
fine precision at low cost. Then (section~\ref{sec:analytic_comput})
we will introduce another model where~$V$ is a piecewise constant
which will allow us to go further in the analytical
computations. Comparison between the two will strengthen the
interpretation of resonnant tunnelling in terms of avoided
degeneracies in the spectrum and will show that the Fabry-P\'erot
effect provides a correct interpretation. The two models encapsulate the same
physics  and, indeed,  their comparison may help to support this intuition. 
What seems more  interesting for pedagogical sake
is to consider the opposite way : starting with the intuition that
the two models are physically qualitatively equivalent, the very different
methods to study them (numerics, level dynamics, Husimi distribution for the smooth
case and analytical computations, semiclassics, transfer matrices for
the other one) appear to be complementary and  mutually supporting. 

All along this paper, we will use mainly physical concepts and mathematical
tools that correspond to a undergraduate course on quantum 
physics. 
However, the accumulation of different techniques 
(specially the somehow lengthy computations of section 3.2,
 a reasonable skill in asymptotic expansions,
 the numerical implementation and the notion of quasiclassical states) 
are more accessible once the very first notions of quantum physics
are well-understood or during a special 
supervised training course.

\section{Three-well polynomial potential }\label{sec:q6pot}
\subsection{Presentation}

The simplest smooth symmetric one-dimensional bounded potential with resonance is given by
\begin{equation}\label{eq:ham_3puits_mou}
V(q)=(q^2-a^2)^2(q^2-b^2),
\end{equation}
where $a$ and $b$ are real. For a suitable choice of the
parameters $(a,b)$, with $b>a$, the potential exhibits two symmetric
external wells separated by a third deeper one (Fig.~\ref{fig:3puits_mou_pot_spectre}a). In the following, we will
systematically work with a non relativistic particle whose mass is taken to unity. The quantum spectrum is purely discrete, bounded
from below  and
the eigenstate~$\ket{\psi}$ of energy~$E$ is given by the
spectral equation
\begin{equation}
\hat{H}\ket{\psi}=E\ket{\psi}
\end{equation}
where the Hamiltonian is~$\hat{H}=H(\hat{p},\hat{q})$ where~$\hat{p},\hat{q}$ are the momentum and position  operators and~$H$ the classical Hamiltonian
\begin{equation}\label{eq:hamiltonian}
  H(p,q) = \frac{p^2}{2} + V(q)\;.
\end{equation}

\subsection{The spectrum}

When dealing with a polynomial Hamiltonian, a
convenient way to perform the numerical
computations is to write directly the canonical operators
in the harmonic oscillator eigenbasis
\cite{Korsch/Gluck02a}
\begin{equation}
\hat{q} = \sqrt{\frac{\hbar}{2}}
\begin{pmatrix}
0 & \sqrt{1} & 0 & 0 & \cdots \\
\sqrt{1} & 0 & \sqrt{2} & 0 & \cdots \\
0 & \sqrt{2} & 0 & \sqrt{3} & \cdots \\
0 & 0 & \sqrt{3} & 0 & \cdots \\
\vdots & \vdots & \vdots & \vdots & \ddots
\end{pmatrix}, \hat{p} = \imat \sqrt{\frac{\hbar}{2}}
\begin{pmatrix}
0 & -\sqrt{1} & 0 & 0 & \cdots \\
\sqrt{1} & 0 & -\sqrt{2} & 0 & \cdots \\
0 & \sqrt{2} & 0 & -\sqrt{3} & \cdots \\
0 & 0 & \sqrt{3} & 0 & \cdots \\
\vdots & \vdots & \vdots & \vdots & \ddots
\end{pmatrix}.
\end{equation}
The Hamiltonian \eqref{eq:hamiltonian} is now written in terms of power of these infinite matrices. In practice, one needs to truncate them. First, one computes numerically the products of the truncated matrices and then one sums each terms. One can now diagonalize the Hamiltonian to get the eigenvalues. Numerically, we must finally check that the
  truncation of these infinite matrices does not affect the results up
  to the required precision.

\begin{figure}[!ht]
\center \includegraphics[height=0.6\textwidth,width=1.0\textwidth]{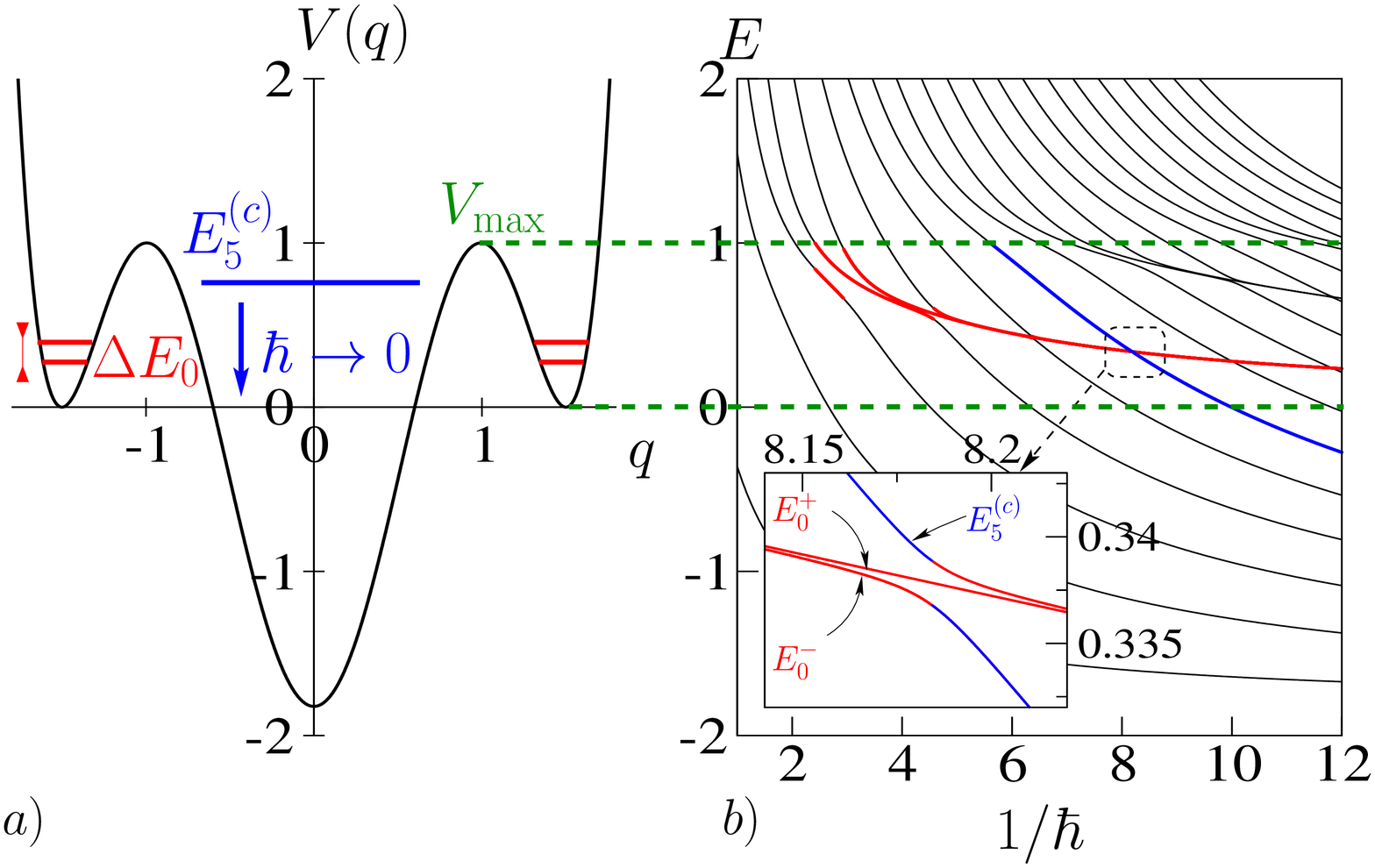}
\caption{\label{fig:3puits_mou_pot_spectre} a)~The three-well
  potential \eqref{eq:ham_3puits_mou} for $a=1.5$ and $b=0.6$. b)
  The first $20$ energy levels versus $1/\hbar$ (black lines in figure b). The dashed
  green lines delimit the energy range where the doublets $\Delta E_n$
  lie. We have also sketched the doublet $\Delta E_{0}$ (in red)
  and the level~$E^{(c)}_{5}$ (in blue)
  associated to the central well.  The inset in b) shows the
  avoided crossing between the three levels~$E_0^+$, $E_0^-$
  and~$E_{5}^{(c)}$ that occurs near~$1/\hbar\simeq8.18$.
  }
\end{figure}

The eigenenergies are plotted as a
function of the Planck's constant $\hbar$ in
figure~\ref{fig:3puits_mou_pot_spectre}b). Even though it is a standard procedure when one studies quantum phenomena in the semiclassical limit, it is worth to remind that treating the Planck constant as a varying parameter has to be understood as a substitute for varying the physical parameters of the system (for instance, through a rescaling of the quantities including $\hbar$  in order to work with dimensionless parameters). Since~$V(-q)=V(q)$,
the eigenstates can be classified according to their parity ($\pm$).
Moreover, for small enough $\hbar$, the part of the spectrum below the local maximum~$V_{\mathrm{max}}$ 
can be divided in two families: first, a family of doublets~$(E^+_n,E^-_n)$ that remain positive and below~$V_{\mathrm{max}}$
for any value of~$\hbar$ (like the red lines in fig.~\ref{fig:3puits_mou_pot_spectre}b))  and second, a family of energies
 that eventually become negative for small enough~$\hbar$ (like the blue line in fig.~\ref{fig:3puits_mou_pot_spectre}b)).
Except near the special values of~$\hbar$ where the two families almost intersect (these crossings will be discussed in the next section) the first family
corresponds to states localised in the lateral wells with~$E^+_n\simeq E^-_n\sim\hbar\omega_l(n+1/2)$, with~ integer~$n$ and~$\omega_l=\sqrt{V''(a)}$ the harmonic frequency in the bottom of the lateral wells  :  like in the double-well case, the difference~$\Delta E_n\DEF|E_n^+-E_n^-|$
will be our tunnelling splittings between symmetric and antisymmetric states. 
The energies that make up the second family do not form doublets but rather correspond to states localised in the central well.
In the semiclassical limit these energies behave 
as~$E_m^{(c)}=V(0)+\hbar\omega_{c}(m+1/2)+\mathrm{O}(\hbar^2)$  with integer~$m$ 
and~$\omega_c=\sqrt{V''(0)}$ the harmonic frequency  in the bottom of the central well.

\subsection{Avoided crossings in the spectra}
 In the absence of singularities \cite{Cohen/Kuharetz93a}, the spectrum has no
  degeneracies.  When varying a control parameter ($\hbar$ in our
  case), what appears to be a crossing between energies is actually an
  avoided crossing if one look at a sufficiently high resolution (see
  the insert in figure~\ref{fig:3puits_mou_pot_spectre}b)).  More
  precisely, two energy levels having the same parity seem to repel
  one from the other and a substitution between one component of the
  doublet ($E^-_0$) and $E_{5}^{(c)}$ occurs when~$1/\hbar$ increases
  from~8.14 to~8.24. This scenario can be reproduced using a three
-level model (\S~\ref{sec:mat_model}) and can be followed using a
phase-space representation of the corresponding wave functions
(\S~\ref{subsec:husimi}).

\begin{figure}[!ht]
\center \includegraphics[height=0.8\textwidth,width=1.0\textwidth]{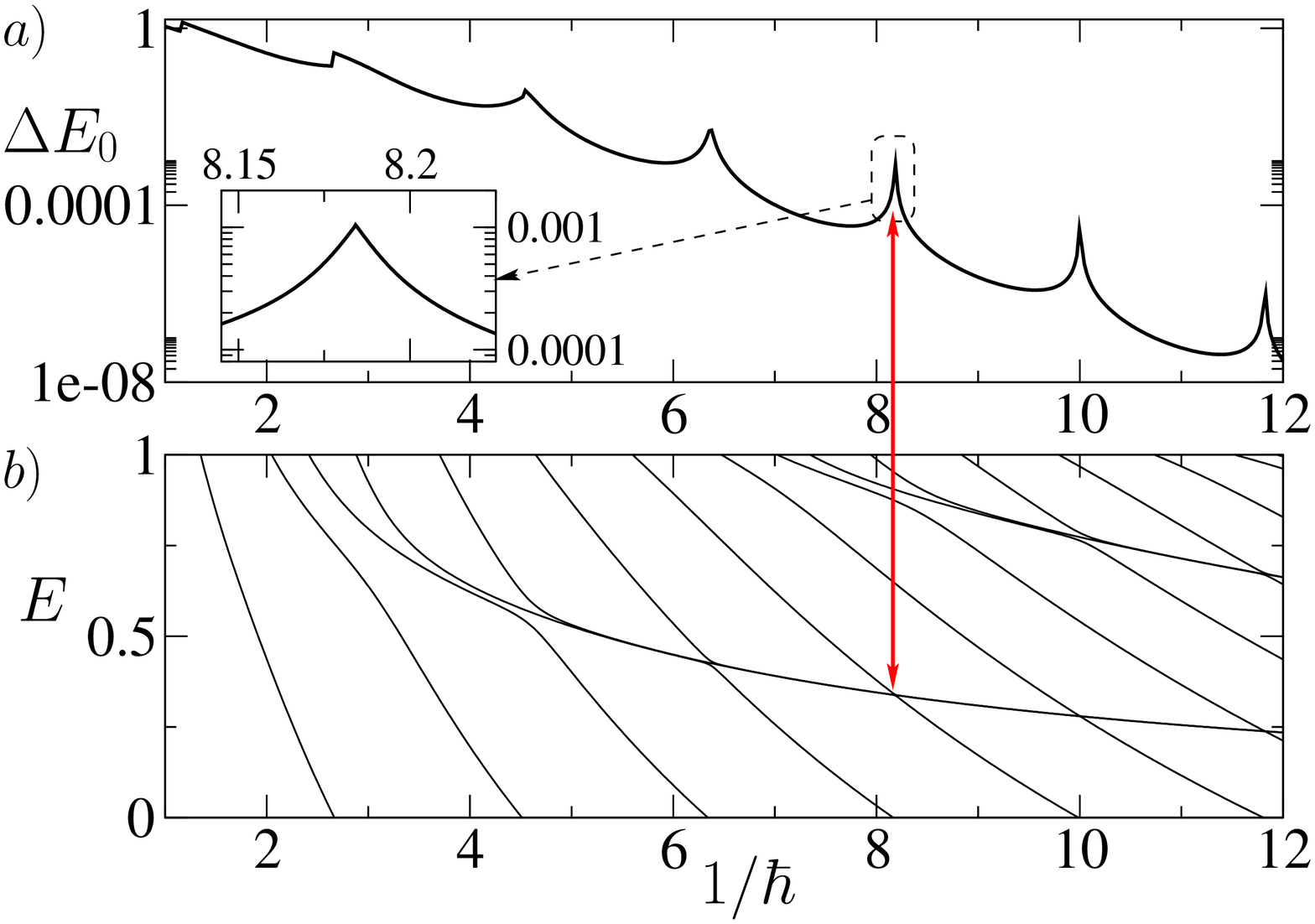}
\caption{\label{fig:3puits_mou_split_spectre} The splitting $\Delta
  E_0$ is plotted in a) in a semilogarithmic
  scale versus $1/\hbar$ for the three-well potential
  \eqref{eq:ham_3puits_mou} with $a=1.5$ and $b=0.6$. The figure b)
  exhibits the associated spectra versus $1/\hbar$. The double arrowed
  red line indicates the correspondance between avoided crossing in
  the spectra and resonances in the splitting. The resonance at $\hbar
  \simeq 1/8.18$ has been magnified in the insert in figure a).}
\end{figure}

Let us note that near such avoided crossings, the definition of the doublet becomes ambiguous: 
whatever criterion we retain to discriminate between the levels $E_n^\pm$ and $E_m^{(c)}$
(for instance the two nearest energies or those states selected from their overlap with a coherent state
localised in one of the three wells) a discontinuous switch will occur and the level repulsion originates
a large increase, by several of magnitude, of the tunnelling splittings $\Delta E_n$. Those are the resonances
seen from the energy levels point of view. The word ``resonance'' is then to be taken in the usual sense: the strong variation of~$\Delta E$ within a small range of variation of one parameter ($1/\hbar$) is directly related
to the coincidence between frequencies, namely the frequencies of the Rabi oscillations 
between the wells. From then on, we will define $\Delta E_n$ in selecting the energies involved as the largest overlap between the eigenstates and a coherent state localised in one of the external well. The splitting will thus be computed numerically following this criterion, leading to the figures \ref{fig:3puits_mou_split_spectre}a) and \ref{fig:3puits_carre_split}. As seen in figure~\ref{fig:3puits_mou_split_spectre}a), 
the simple exponential behaviour \eqref{eq:deltaE}
breaks down when the doublet is mixed with a third level and some resonant spikes appear. Moreover,
these peaks are not differentiable
at the values of~$\hbar$ corresponding to the discontinuous
switches. As $\Delta E$ increases close to a
resonance, the tunnelling period $T=\hbar/\Delta E$ becomes smaller: a
quantum state initially prepared in one of the external wells will
then oscillate more rapidly from this well to its symmetric. For some
particular values of the parameters, that is to say when an isolated
level comes to perturb the doublet, a middle well separating two
symmetric wells can thus enhanced significantly the
tunnelling\footnote{In more complicated systems which are beyond the
  scope of this paper, one can observe crossings in the spectra and
  then the splitting would exhibit antipeaks with infinite height in a
  semilogarithmic scale. The tunnelling would thus be completely
  destroyed.}. In section \ref{sec:analytic_comput} we will derivate
almost analytical formulas for the splitting and the height of the
peaks.

\subsection{Husimi representation}\label{subsec:husimi}

\begin{figure}[!htb]
\center \includegraphics[height=1.2\textwidth,width=1.0\textwidth]{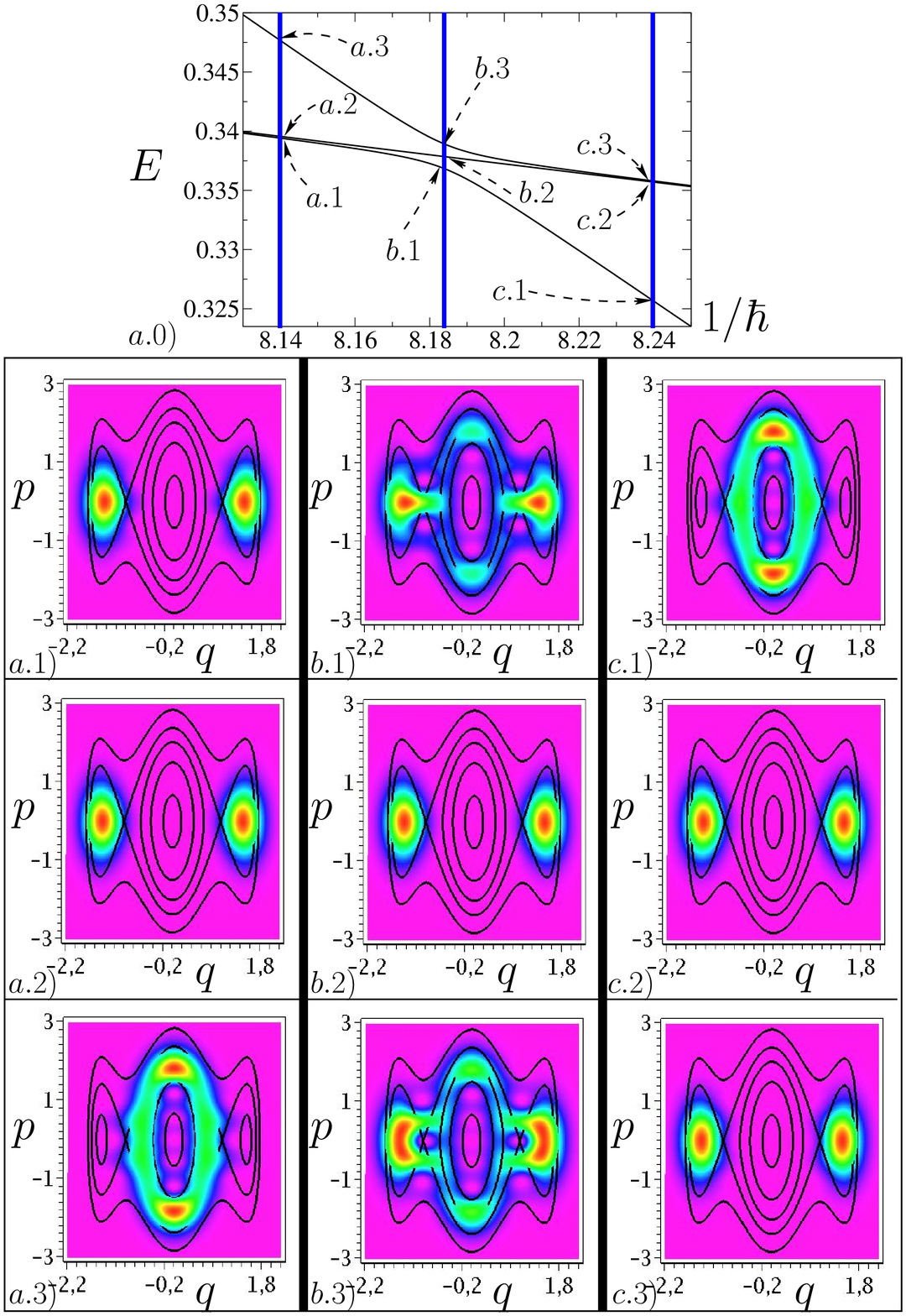}
\caption{\label{fig:husimi} For the three-well polynomial potential
  \eqref{eq:ham_3puits_mou} with $a=1.5$ and $b=0.6$, we have plotted
  the Husimi distribution for the three levels involved in the avoided
  crossing around $\hbar \simeq 1/8.18$ (see also insert in figure
  \ref{fig:3puits_mou_pot_spectre}a). The vertical blue lines in
  respectively a,b,c.0) defines the values of the Planck's constant
  (resp. $1/\hbar \simeq 8.14, 8.184, 8.24$) and shows the relative
  position of the energies in the spectra (black lines) where the
  dashed arrows point out the energy levels related to each Husimi
  distribution depicted in a,b,c.1,2,3). The classical phase space is
  also superimposed on the Husimi plots.}
\end{figure}
It is very illuminating to see the resonance at work using a
  phase-space representation of the eigenfunctions involved in the
  triplet~$E_n^\pm,E_m^{(c)}$ near an avoided crossing.
Instead of computing eigenstates in either
the $q-$ or $p-$representation, we will choose the Husimi
distribution $\psi(p,q)$
which mixes both space and momentum variables and gives the density
probability to find a quantum state in the phase space. The Husimi
distribution is defined as the square modulus of the projection of the
state $\ket{\psi}$ onto a coherent state $\ket{\alpha}$~\cite[complement G$_{\textrm{V}}$]{Cohen/Diu/Laloe97a}. One can use
the basis $\ket{n}$ of the harmonic oscillator in order to write the
coherent state as follows
\begin{equation}
\ket{\alpha} = \EXP{-|\alpha|^2/2}\sum_{n=0}^{\infty}{\frac{\alpha^2}{\sqrt{n!}}\ket{n}},
\end{equation}
where $\alpha \DEF (p + \imat q)/\sqrt{2\hbar}$, then 
\begin{equation}
\psi(p,q) \DEF \left| \braket{\alpha}{\psi} \right|^2 = \EXP{-|\alpha|^2} \left| \sum_{n=0}^{\infty}{\frac{\alpha^2}{\sqrt{n!}}\braket{\psi}{n}} \right|^2.
\end{equation}
The Husimi distribution $\psi(p,q)$ is definite
positive. The coherent state $\ket{\alpha}$ is a Gaussian 
wavepacket built such that it saturates the uncertainty inequalities i.e. $\Delta p \Delta q = \hbar/2$. A coherent state, also called a quasi-classical state, can be understood as the \textit{most classical} quantum state since it can be shown that the expectation values of the quantum observables follow 
the same evolution as the corresponding classical ones at the leading order of $\hbar$. Physically, the square of the scalar product of the eigenstate $\ket{\psi}$ and a coherent state can be interpreted as the density probability to find the quantum state in a cell of area $\hbar$ centered at the momentum $p$ and the position $q$ in a coarse-grained phase space. The Husimi distribution appears thus to be an appropriate picture to study quantum dynamics in the semiclassical regime~\cite{Novaes03a}. Far from a resonance (see figure \ref{fig:husimi}a.0), the
symmetric and antisymmetric eigenstates related to the doublet with
the energies $(E_0^+,E_0^-)$ are mainly localised in the external
wells while the isolated level is clearly well-delimited in the middle
well as shown in figures \ref{fig:husimi}a.1,2,3). When $1/\hbar$ increases
  from~8.15 to~8.25, we can follow the exchange between the state corresponding to $E_{5}^{(c)}$
and the state corresponding to~$E_0^-$ while the even state remains almost unaltered and mainly
localised in the lateral wells. At the resonance (fig. \ref{fig:husimi}b.1,2,3), the two states with the same parity become intertwined: this is the signature in the phase space of the so-called resonances. For these particular values of the parameters, the odd quantum states leak out through the barriers all over the three wells. We cannot clearly distinguish which odd state is now a part of the doublet and we find the ambiguity concerning the definition of the splitting again.

\section{Almost analytical computations}\label{sec:analytic_comput}
\subsection{Square potential}
\begin{figure}[!ht]
\center \includegraphics[height=0.5\textwidth,width=0.7\textwidth]{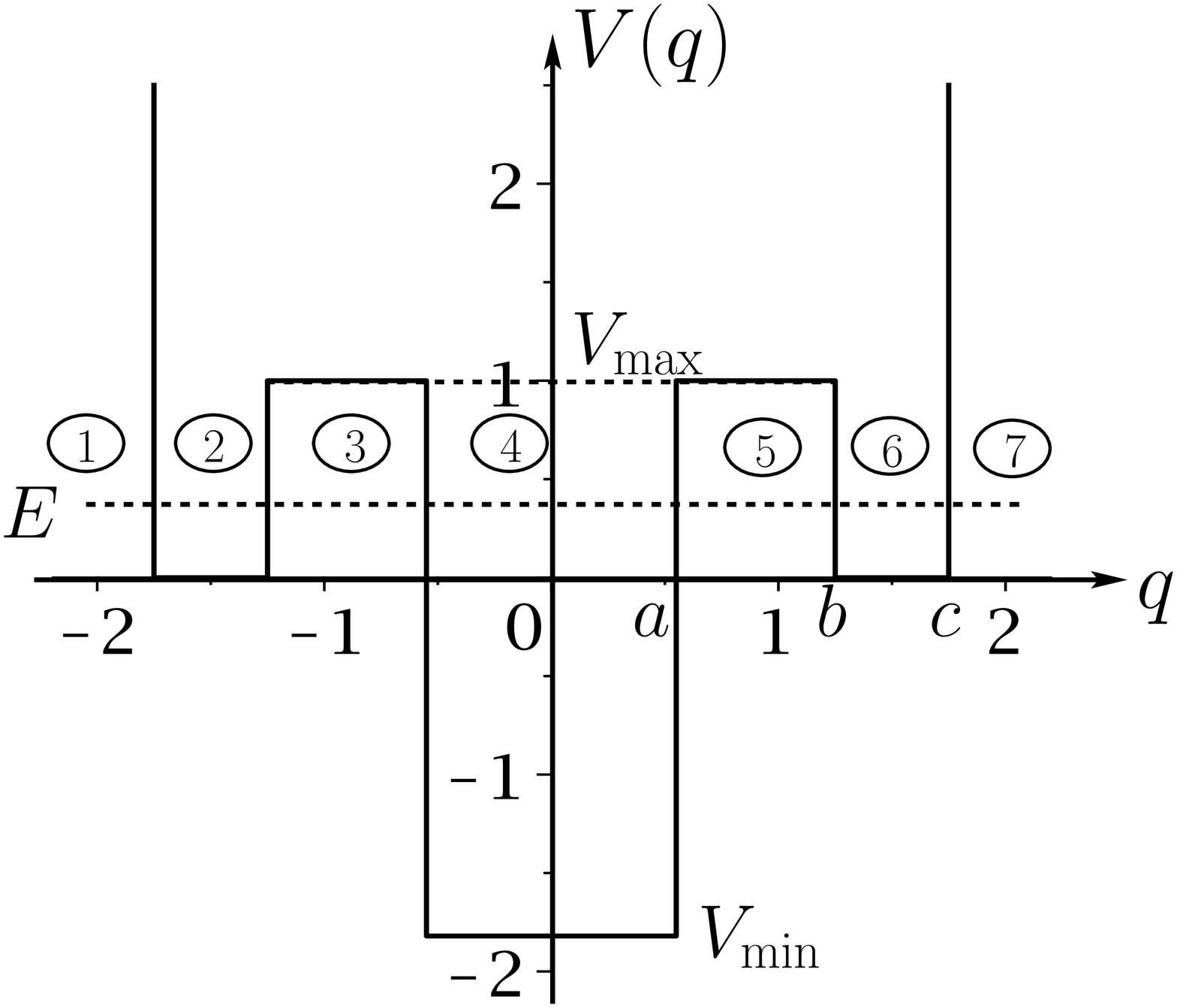}
\caption{\label{fig:3puits_carre} The piecewise three-well square
  potential defined in \eqref{eq:3puits_carre} with the parameters
  $a=0.55$, $b=1.25$, $c=1.75$, $V_{\mathrm{max}}=1.0$ and
  $V_{\mathrm{min}}=-1.82$.}
\end{figure}

To go beyond a simple diagonalization and understand analytically
  the origin of the resonant spikes in the splittings, potential \eqref{eq:ham_3puits_mou}
 is still difficult to handle. The standard instanton procedure of Wick rotating the time~$t\mapsto\imat t$
in order to work with a upside-down potential \cite[chap.~7]{Coleman85a} fails because there remain
a barrier to cross when reversing~$V$ into~$-V$ in
figure~\ref{fig:3puits_mou_pot_spectre}~a). Two of us have recently
 shown how to generalise the instanton method
in the most general $1d$-case  \cite{Ledeunff/Mouchet10a} and how to recover analytically the effect of resonances but 
this method goes far beyond the present elementary complement to a first years lecture on quantum physics.
To deal with a simpler analytically tractable model let us
  introduce a three-well piecewise constant potential that will mimic
  the situation described in the previous section. Therefore we will take (Fig.~\ref{fig:3puits_carre})
  
\begin{equation}\label{eq:3puits_carre}
  V(q) =
  \begin{cases}  
    +\infty \qquad & \text{Region 1: } \qquad \, \, \, \, \, \, q<-c ,\\
    0 \qquad & \text{Region 2: } -c <q< -b ,\\
    V_{\mathrm{max}} \qquad & \text{Region 3: } -b <q< -a ,\\
    V_{\mathrm{min}} \qquad & \text{Region 4: } -a <q< a ,\\
    V_{\mathrm{max}} \qquad & \text{Region 5: }  \, \, \, \, \, \, \, a <q< b ,\\
    0 \qquad & \text{Region 6: } \, \, \, \, \, \, \, b <q< c ,\\
    +\infty \qquad & \text{Region 7: } \, \, \, \, \, \, \, c <q.
  \end{cases}
\end{equation}
While keeping the essential features of resonances described above, we will then be able to derive
analytical formulas in the semiclassical limit
for the splitting and the height of the resonance peaks.

\subsection{Semiclassical transfer matrix approach}
In the case of one-dimensional systems, a convenient way to obtain the
eigenfunctions and eigenenergies is to use transfer matrix formalism
\cite{Walker/Gathwright94a} \cite[Appendix~A]{Azbel83a}. The transfer
matrices connect the amplitudes of the wavefunctions in the different
regions of the potential. As we want to compute tunnelling splittings,
we will pay particular attention to energies such that
$0<E<V_{\mathrm{max}}$. We will use this approach to get the equations
which give the energy levels and secondly to compute the splitting in
the semiclassical regime.

\subsubsection{Energy quantization}
With potential~\eqref{eq:3puits_carre}, the general solution~$\psi(q)$ of the time-independent  Schr\"{o}dinger's equation at
 energy $E$,
  $-\hbar^{2}\psi''(q)/2 + V(q)\psi(q) = E\psi(q)$,
(the prime stands for the derivative) has the form:
\begin{equation}
\begin{cases}
  \text{Region 1:} &\psi_{1}(q) = 0, \\
  \text{Region 2:} &\psi_{2}(q) = A_{2}\EXP{\imat k_0q} + B_{2}\EXP{-\imat k_0q}, \\
  \text{Region 3:} &\psi_{3}(q) = A_{3}\EXP{k_{\mathrm{max}}q} + B_{3}\EXP{-k_{\mathrm{max}}q}, \\
  \text{Region 4:} &\psi_{4}(q) = A_{4}\EXP{\imat k_{\mathrm{min}}q} + B_{4}\EXP{-\imat k_{\mathrm{min}}q}, \\
  \text{Region 5:} &\psi_{5}(q) = A_{5}\EXP{k_{\mathrm{max}}q} + B_{5}\EXP{-k_{\mathrm{max}}q}, \\
  \text{Region 6:} &\psi_{6}(q) = A_{6}\EXP{\imat k_0q} + B_{6}\EXP{-\imat k_0q}, \\
  \text{Region 7:} &\psi_{7}(q) = 0,
\end{cases}
\end{equation}
where
\begin{equation}\label{eq:w_num}
k_0 = \frac{\sqrt{2E}}{\hbar}, \quad k_{\mathrm{min}} = \frac{\sqrt{2(E-V_{\mathrm{min}})}}{\hbar}, \quad k_{\mathrm{max}} = \frac{\sqrt{2(V_{\mathrm{max}}-E)}}{\hbar}.
\end{equation}
The amplitudes $(A_i,B_i)$ are \textit{a priori} complex numbers. As the energy $E>0$, the wavefunctions $\psi_{2}(q)$,
$\psi_{4}(q)$ and $\psi_{6}(q)$ are linear combinations of two plane
waves travelling in opposite directions, while
in the regions where the energy is lower than the barrier,
$\psi_{3}(q)$ and $\psi_{5}(q)$ are combinations of real exponentials.
Though the potential $V(q)$ is not continuous, one still requires the
eigenfunctions and its derivative to be smooth everywhere and
especially at each frontier between two regions, i.e. $q=\{\pm a,\pm
b,\pm c \}$. It leads to the following equalities
\begin{equation}
\begin{cases}\label{eq:continuity_3puits_carre}
  \psi_{2}(-c) = 0, \\
  \psi_{2}(-b) = \psi_{3}(-b), \qquad  \, \psi'_{2}(-b) = \psi'_{3}(-b), \\
  \psi_{3}(-a) = \psi_{4}(-a), \qquad \psi'_{3}(-a) = \psi'_{4}(-a), \\
  \psi_{4}(a) = \psi_{5}(a), \qquad \, \, \, \, \, \, \, \, \, \psi'_{4}(a) = \psi'_{5}(a), \\
  \psi_{5}(b) = \psi_{6}(b), \qquad \, \, \, \, \, \, \, \, \, \, \psi'_{5}(b) = \psi'_{6}(b), \\
  \psi_{6}(c) = 0. \\
\end{cases}
\end{equation}
Using these continuity equations, one can express the
amplitudes $(A_i,B_i)$ in terms of any others $(A_{j},B_{j})$ through
the relation
\begin{equation}
\begin{pmatrix}
A_i \\
B_i
\end{pmatrix}= \grT(i,j)
\begin{pmatrix}
A_{j} \\
B_{j}
\end{pmatrix}, \qquad \grT(i,j) =
\begin{pmatrix}
\grT_{11}(i,j) & \grT_{12}(i,j) \\
\grT_{21}(i,j) & \grT_{22}(i,j)
\end{pmatrix},
\end{equation}
where $\grT(i,j)$ is the transfer matrix from the region $j$ to $i$. Each transfer matrix can be written using three elementary matrices
\begin{equation}
\begin{cases}
&\gru_{\alpha}(k) =
\begin{pmatrix}
\EXP{\imat k \alpha} & 0 \\
0 & \EXP{-\imat k \alpha}
\end{pmatrix}, \quad 
\grv_{\alpha}(k) =
\begin{pmatrix}
\EXP{-k \alpha} & 0 \\
0 & \EXP{k \alpha}
\end{pmatrix}, \\
&\grw(k,\tilde{k}) = \displaystyle{\frac{1}{2}}
\begin{pmatrix}
1-\imat k/\tilde{k} & 1+\imat k/\tilde{k} \\
1+\imat k/\tilde{k} & 1-\imat k/\tilde{k}
\end{pmatrix},
\end{cases}
\end{equation}
where $k$, $\tilde{k}$ and $\alpha$ are real. The matrices~$\gru$ (resp.~$\grv$)  correspond to 
the propagation in a classicaly allowed region ($E>V$) (resp. forbidden region, $E<V$); 
the matrices~$\grw$ correspond
to the continuity equations at one step. We have
\begin{equation}
\gru_{-\alpha}(k) = \big[\gru_{\alpha}(k)\big]^{-1}, \qquad \gru_{\alpha}(k) \gru_{\beta}(k) = \gru_{\alpha + \beta}(k).
\end{equation}
and
$\grw$ fulfills the condition $[\grw(k,\tilde{k})]^{-1}
= \grw^*(\tilde{k},k)$, for real $k$ and $\tilde{k}$, where $\grw^*$  denotes the complex conjugate of $\grw$. The
transfer matrices are
given by
\begin{eqnarray}
\grT(2,3) &=& \gru_{b}(k_0)\grw(k_{\mathrm{max}},k_{0})\grv_{b}(k_{\mathrm{max}}), \\
\grT(3,4) &=& \grv_{-a}(k_{\mathrm{max}})\grw^{*}(k_{\mathrm{min}},k_{\mathrm{max}})\gru_{-a}(k_{\mathrm{min}}), \\
\grT(4,5) &=& \gru_{-a}(k_{\mathrm{min}})\grw(k_{\mathrm{max}},k_{\mathrm{min}})\grv_{-a}(k_{\mathrm{max}}), \\
\grT(5,6) &=& \grv_{b}(k_{\mathrm{max}})\grw^{*}(k_{0},k_{\mathrm{max}})\gru_{b}(k_{0}).
\end{eqnarray}
Then, the relation between $(A_2,B_2)$ and $(A_4,B_4)$ are given by the transfer matrix : 
\begin{eqnarray}\label{eq:trans_mat_l}
\grT(2,4) \hspace{-0.2cm} &=& \hspace{-0.2cm} \grT(2,3) \grT(3,4) \notag \\
&=& \hspace{-0.2cm} \gru_{b}(k_0)\grw(k_{\mathrm{max}},k_{0})\grv_{b}(k_{\mathrm{max}}) \bigg [ \gru_{a}(k_{\mathrm{min}})\grw(k_{\mathrm{max}},k_{\mathrm{min}})\grv_{a}(k_{\mathrm{max}}) \bigg ]^{-1}.
\end{eqnarray}
A straightforward computation shows that the elements of the matrix
transfer through a barrier have to be such that $\grT_{22} =
(\grT_{11})^{*}$ and $\grT_{21} = (\grT_{12})^{*}$. 
Since the potential is symmetric, the matrix $\grT(2,4)$ contains already all the information
about the system. Indeed, noting that
\begin{eqnarray}\label{eq:trans_mat_r}
\grT(6,4) \hspace{-0.2cm} &=& \hspace{-0.2cm} [\grT(4,6)]^{-1} = [\grT(5,6)]^{-1} [\grT(4,5)]^{-1} \\
&=& \hspace{-0.2cm} \gru_{-b}(k_0)\grw(k_{\mathrm{max}},k_{0})\grv_{-b}(k_{\mathrm{max}}) \bigg [ \gru_{-a}(k_{\mathrm{min}})\grw(k_{\mathrm{max}},k_{\mathrm{min}})\grv_{-a}(k_{\mathrm{max}}) \bigg ]^{-1}, \notag
\end{eqnarray}
and performing the symmetry-axis transformation $(a,b,c)\rightarrow
(-a,-b,-c)$, the expression \eqref{eq:trans_mat_r} is nothing but
$\grT(2,4)$. We distinguish the parity of the eigenstates with the
condition $\psi_4'(0)=0$ for the even states while the odd ones
fulfill $\psi_4(0)=0$. Using the continuity condition at $q=-c$, it
leads to the algebraic system
\begin{equation}
\begin{pmatrix}
A_2 \\
-A_2\EXP{-2\imat k_0 c}
\end{pmatrix}=
\begin{pmatrix}
\grT_{11}(2,4) & \grT_{12}(2,4) \\
(\grT_{12}(2,4))^{*} & (\grT_{11}(2,4))^{*}
\end{pmatrix}
\begin{pmatrix}
A_4 \\
\pm A_4
\end{pmatrix},
\end{equation}
where $\pm$ stands for the parity of the
eigenstate.  The transcendantal equations for the energy
levels are thus obtained easily by manipulating the system and
eliminating the amplitudes from the previous equalities:
\begin{eqnarray}
& \re{\left[ (\grT_{11}(2,4) + \grT_{12}(2,4))\EXP{-\imat k_0 c} \right]} = 0& \quad \text{: even states,} \\
& \im{\left[ (\grT_{11}(2,4) - \grT_{12}(2,4))\EXP{-\imat k_0 c} \right]} = 0& \quad \text{: odd states.}
\end{eqnarray}
These conditions can be written in a more tractable form for the even states
\begin{equation}\label{eq:3puits_mou_nrj_e}
D_{+}(E) \DEF F_{+}(E) + G_{+}(E)\EXP{-2k_{\mathrm{max}}(b-a)} = 0,
\end{equation}
and the odd states
\begin{equation}\label{eq:3puits_mou_nrj_o}
D_{-}(E) \DEF F_{-}(E) - G_{-}(E)\EXP{-2k_{\mathrm{max}}(b-a)} = 0,
\end{equation}
with
\begin{eqnarray}
&& F_{+}(E) \DEF \bigg(k_0\cos[k_0(c-b)] + k_{\mathrm{max}}\sin[k_0(c-b)] \bigg) \label{eq:Fe} \\
&& \qquad \qquad \qquad \qquad \qquad \qquad \qquad \times \bigg( k_{\mathrm{max}} \cos(k_{\mathrm{min}}a) - k_{\mathrm{min}}\sin(k_{\mathrm{min}}a) \bigg),\notag \\
&& G_{+}(E) \DEF \bigg(k_0\cos[k_0(c-b)] - k_{\mathrm{max}}\sin[k_0(c-b)] \bigg) \label{eq:Ge} \\
&& \qquad \qquad \qquad \qquad \qquad \qquad \qquad \times \bigg( k_{\mathrm{max}} \cos(k_{\mathrm{min}}a) + k_{\mathrm{min}}\sin(k_{\mathrm{min}}a) \bigg), \notag \\
&& F_{-}(E) \DEF \bigg(k_0\cos[k_0(c-b)] + k_{\mathrm{max}}\sin[k_0(c-b)] \bigg) \label{eq:Fo} \\
&& \qquad \qquad \qquad \qquad \qquad \qquad \qquad \times \bigg( k_{\mathrm{min}}\cos(k_{\mathrm{min}}a) + k_{\mathrm{max}}\sin(k_{\mathrm{min}}a) \bigg), \notag \\
&& G_{-}(E) \DEF \bigg(k_0\cos[k_0(c-b)] - k_{\mathrm{max}}\sin[k_0(c-b)] \bigg) \label{eq:Go} \\
&& \qquad \qquad \qquad \qquad \qquad \qquad \qquad \times \bigg( k_{\mathrm{min}}\cos(k_{\mathrm{min}}a) - k_{\mathrm{max}}\sin(k_{\mathrm{min}}a) \bigg). \notag 
\end{eqnarray}
The zeroes of $D_\pm(E)$ give the exact discrete energies $E=E_n^{\pm}$ for the three-well square potential.

In the semiclassical limit, the height of the barrier
is much larger than the energies and the term of order one in the sums
\eqref{eq:3puits_mou_nrj_e} and \eqref{eq:3puits_mou_nrj_o} dominate the decreasing exponential, so the energies can be
approximated keeping only the first term
\begin{eqnarray}
D_{+}(E) \underset{\hbar \rightarrow 0}{\sim} F_{+}(E) = 0, \label{eq:3puits_mou_nrj_e_approx} \\
D_{-}(E) \underset{\hbar \rightarrow 0}{\sim} F_{-}(E) = 0. \label{eq:3puits_mou_nrj_o_approx}
\end{eqnarray}
Within this approximation, tunnelling is neglected. There are two possible ways to cancel \eqref{eq:3puits_mou_nrj_e_approx} or \eqref{eq:3puits_mou_nrj_o_approx} as the functions $F_{+}(E)$ and $F_{-}(E)$ are written as a product of two functions
\begin{eqnarray}
F_{+}(E) &=& F_{\goe}(E) \times F_{\gom ,+}(E), \label{eq:Fe_def}\\
F_{-}(E) &=& F_{\goe}(E) \times F_{\gom ,-}(E), \label{eq:Fo_def}
\end{eqnarray}
with
\begin{equation}
\begin{cases}
F_{\goe}(E) &\DEF \quad k_0\cos[k_0(c-b)] + k_{\mathrm{max}}\sin[k_0(c-b)], \\
F_{\gom ,+}(E) &\DEF \quad k_{\mathrm{max}} \cos(k_{\mathrm{min}}a) - k_{\mathrm{min}}\sin(k_{\mathrm{min}}a), \\
F_{\gom ,-}(E) &\DEF \quad k_{\mathrm{min}}\cos(k_{\mathrm{min}}a) + k_{\mathrm{max}}\sin(k_{\mathrm{min}}a).
\end{cases}
\end{equation}

The factor $F_{\goe}(E)$ appears in both cases (even and odd) and
corresponds to the quantization condition in the lateral wells in the limit $b-a\rightarrow +\infty$ and provide the average position $e_n$ of the doublets. The
second factors $F_{\gom ,\pm}(E)$ in \eqref{eq:Fe_def} and
\eqref{eq:Fo_def} are nothing but the transcendantal equations which
lead respectively to the even and odd energy levels $e^{\pm}_{\gom}$
of the finite square well \cite[chap.~6, \S~8]{Merzbacher70a} and
approximate the energies related to the central well of
\eqref{eq:3puits_carre}. It is clear that, in the limit where we do
not care about tunnelling, the energies associated to the external
wells are degenerate as both quantization conditions
\eqref{eq:3puits_mou_nrj_e_approx} and
\eqref{eq:3puits_mou_nrj_o_approx} share the same first
factor. Avoided crossings also disappear as the two factors of
$F_{+}(E)$ (resp. $F_{-}(E)$) in \eqref{eq:Fe_def}
(resp. \eqref{eq:Fo_def}) vanish simultaneously for some particular
values\footnote{For these particular values of $\hbar$, some zeros of
  the conditions \eqref{eq:3puits_mou_nrj_e_approx} and
  \eqref{eq:3puits_mou_nrj_o_approx} are of order 2 and give the
  average positions of the actual avoided crossings, otherwise they
  are all of order one.} of $\hbar$.

\subsubsection{Semiclassical formulas for the splitting}
\begin{figure}[!ht]
\center \includegraphics[height=0.6\textwidth,width=0.8\textwidth]{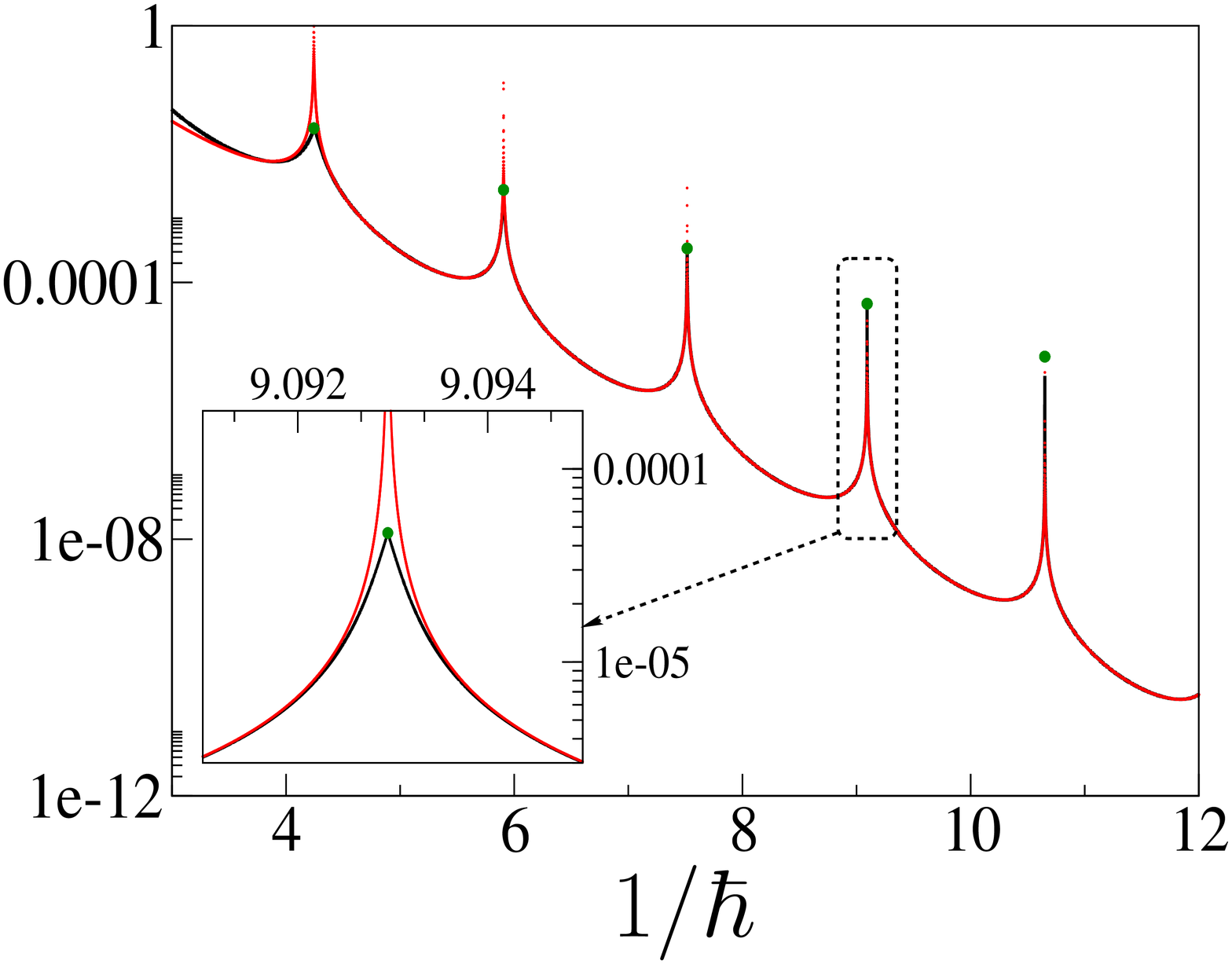}
\caption{\label{fig:3puits_carre_split} The splitting for the ground
  state is plotted in a semilogarithmic scale versus $1/\hbar$ for the
  piecewise constant potential defined in \eqref{eq:3puits_carre} with
  the parameters $a=0.55$, $b=1.25$, $c=1.75$, $V_{\mathrm{max}}=1.0$
  and $V_{\mathrm{min}}=-1.82$. The black line is the exact
  computation obtained by solving numerically the exact quantization
  conditions \eqref{eq:3puits_mou_nrj_e} and
  \eqref{eq:3puits_mou_nrj_o}. The red line corresponds to the
  semiclassical formula \eqref{eq:splitting} and the green dots show
  the height of the resonances computed with \eqref{eq:splitting_e}
  and \eqref{eq:splitting_o}. The insert is a magnification of the
  resonance at $\hbar \simeq 1/9.093$.}
\end{figure}

To recover the exponentially fine
structure of the spectra that caracterise tunnelling, let us expand
the right hand side of the exact quantization
conditions \eqref{eq:3puits_mou_nrj_e} and \eqref{eq:3puits_mou_nrj_o}
around the average value of the doublet that has been determined above $E^{\pm}_n = e_n + \epsilon^{\pm}_n$
\begin{eqnarray}\label{eq:dl_e}
D_{+}(e_n + \epsilon^{+}_{n}) &\simeq & D_{+}(e_n) + \epsilon^{+}_{n}D'_{+}(e_n) + \omat((\epsilon^{+}_{n})^2) = 0, \\
D_{-}(e_n + \epsilon^{-}_{n}) &\simeq & D_{-}(e_n) + \epsilon^{-}_{n}D'_{-}(e_n) + \omat((\epsilon^{-}_{n})^2) = 0,\label{eq:dl_o}
\end{eqnarray}
Keeping only the exponentially dominant terms, we get
\begin{equation}\label{eq:splitting}
\Delta E_n=|E^{-}_n-E^{+}_n|=|\epsilon^{-}_{n} - \epsilon^{+}_{n}| \underset{\hbar \rightarrow 0}{\sim} \left| \frac{G_{-}(e_n)}{F'_{-}(e_n)} + \frac{G_{+}(e_n)}{F'_{+}(e_n)} \right| \EXP{-2k_{\mathrm{max}}(b-a)}
\end{equation}
where $F'_{\pm}(e_n)=F'_{\goe}(e_n)
\times F_{\gom ,\pm}(e_n)$. One recovers on average the exponentially
small behaviour of the splitting. Noting that $p=\hbar k$ where $k$ is
equal to the expressions in \eqref{eq:w_num}, the argument of the
exponential is nothing but the classical action under the two barriers
of the potential
\begin{equation}
\int_{-b}^{-a}{p\dmat q} + \int_{a}^{b}{p\dmat q} = 2\int_{a}^{b}{p\dmat q} = 2\hbar k_{\mathrm{max}}(b-a).
\end{equation}
While $F'_{\goe}(e_n)\neq0$, one denominator in \eqref{eq:splitting} may still vanish 
when  $F_{\gom ,\pm}(e_n)=0$ which provides a condition for determining the resonances. To fix the divergences
in \eqref{eq:splitting} and get a finite value of the height of the spikes, we have
to go to the next order in the Taylor expansions \eqref{eq:dl_e} and
\eqref{eq:dl_o}. We distinguish both cases depending on
whether the third level is even or odd. If the resonance is due to an
even level $e^{+}_{\gom}$ then $F'_{+}(e_n) = 0$ and we need to expand
$D_{+}(e_n+\epsilon^{+}_{n})$ to the second order. The splitting is of
order $\omat[\EXP{-2k_{\mathrm{max}}(b-a)}]$ but it is no longer true close to a resonance where it becomes much larger because of
the prefactor which is no longer of order one. The dominant terms are
thus now of order $\omat[(\epsilon^{+}_n)^2]$ and $\omat[\EXP{-2k_{\mathrm{max}}(b-a)}]$ while it is still sufficient to keep only the first order term for the odd
condition $D_{-}(e_n+\epsilon^{-}_{n})$, as previously. In case an odd
energy level is responsible for a divergence in \eqref{eq:splitting},
we proceed similarly with the odd condition
$D_{-}(e_n+\epsilon^{-}_{n})$. Collecting all the results we get
\begin{eqnarray}\label{eq:splitting_e}
\Delta E_{n,+} &\underset{\hbar \rightarrow 0}{\sim}& \left| \frac{G_{-}(e_n)}{F'_{-}(e_n)}\EXP{-2k_{\mathrm{max}}(b-a)} - \sqrt{\frac{-2G_{+}(e_n)}{F''_{+}(e_n)}}\EXP{-k_{\mathrm{max}}(b-a)} \right| , \notag \\
 &\underset{\hbar \rightarrow 0}{\sim}& \left| \sqrt{\frac{-2G_{+}(e_n)}{F''_{+}(e_n)}} \right|\EXP{-k_{\mathrm{max}}(b-a)},
\end{eqnarray}
and
\begin{eqnarray}\label{eq:splitting_o}
\Delta E_{n,-} &\underset{\hbar \rightarrow 0}{\sim}& \left| \sqrt{\frac{2G_{-}(e_n)}{F''_{-}(e_n)}}\EXP{-k_{\mathrm{max}}(b-a)} + \frac{G_{+}(e_n)}{F'_{+}(e_n)}\EXP{-2k_{\mathrm{max}}(b-a)} \right| , \notag \\
 &\underset{\hbar \rightarrow 0}{\sim}& \left| \sqrt{\frac{2G_{-}(e_n)}{F''_{-}(e_n)}} \right|\EXP{-k_{\mathrm{max}}(b-a)},
\end{eqnarray}
where $\Delta E_{n,+}$ (resp. $\Delta E_{n,-}$) is the height of a
resonance caused by an even (resp. odd) level from the
central well. The denominators $F''_{\pm}(e_n)$ can be again simplified and written as the product $2F'_{\goe}(e_n) \times
F'_{\gom,\pm}(e_n)$ when the level $e^{\pm}_{\gom}$ is equal to $e_n$
(at the resonance). The argument of the exponential in the splitting
formula \eqref{eq:splitting} far from resonances is twice the argument
in \eqref{eq:splitting_e} and \eqref{eq:splitting_o}. The
semiclassical formula \eqref{eq:splitting} shows a very good agreement
with the exact results as plotted in figure
\ref{fig:3puits_carre_split}. The semiclassical predictions are more
and more accurate for large values of $1/\hbar$ even around the
resonances where the height of the peaks is very well predicted by the
formulas \eqref{eq:splitting_e} and \eqref{eq:splitting_o}.

\subsection{Matrix model}\label{sec:mat_model}
We want to obtain a matrix model of tunnelling in a potential with several wells, as the one of equation (\ref{eq:ham_3puits_mou}). A solution is to represent the Hamiltonian in a basis of "quasi-modes" coupled by tunnelling. Clearly, tunnelling is a wave phenomenon which couples two adjacent wells classically separated by a finite barrier. If there was no tunnelling at all, then each well would have its own independant eigenmodes, approximately given by \textsc{jwkb} ansatz based on the real quantizing classical tori. In the semiclassical limit, tunnelling is a vanishing effect, and these \textsc{jwkb} modes become very good local approximations of the exact solutions of the time independant Schr\"odinger equation, that is, the exact eigenmodes. For this reason, these \textsc{jwkb} ansatz are often called "quasi-mode" in this context. But, if one sticks to real classical tori, then tunnelling is not taken into account by these approximations. On the other hand, one can consider tunnelling as a small coupling between quasi-modes.

For each well of the potential, we hence define a quasi-modes family $\phi_n$, with quasi-energies $E_n$. In the simple case of a symmetric double-well, one has quasi-modes $\phi_n^l$ with energies $E_n^l$, associated with the left well; and quasi-modes $\phi_n^r$ with energies $E_n^r$, associated with the right well. Because of classical symmetry, both are degenerate, that is, $E_n^l=E_n^r$, but tunnelling lifts this degeneracy in the exact spectrum. Close to energy $E_n^l$, the Hamiltonian is well described by a two level system in the basis of the two corresponding quasi-modes. For instance, close to the ground state, one has
\begin{equation}
H\simeq
\begin{pmatrix}
 E_0 & \delta \\ \delta & E_0
\end{pmatrix},
\end{equation}
where $\delta$ gives the coupling strength induced by tunnelling between the two quasi-modes.
Diagonalisation of this matrix gives two eigenstates, an odd one and an even one,
\begin{equation}
\psi_{\pm} = \frac{1}{\sqrt{2}}\left( \phi_0^l\pm\phi_0^r\right),
\end{equation}
associated with two eigenenergies
\begin{equation}
E_{\pm} = E_0\pm \delta.
\end{equation}
In this model, the tunnel splitting is therefore given by
\begin{equation}
\Delta E = 2 \delta.
\end{equation}
Consistency with \textsc{jwkb} analysis and equation (\ref{eq:deltaE}) is obtained by chosing $2\delta = \alpha\hbar e^{-A/\hbar}$.

It is easy to generalise this simple model to a triple well like (\ref{eq:ham_3puits_mou}) by defining three families of quasi-modes : $\phi_n^l$, $\phi_n^c$ and $\phi_n^r$. Now we focus on the situation shown on figure \ref{fig:husimi}. In the language of this model, it is an avoided crossing involving a doublet $\left( \phi_n^l,\phi_n^r\right)$ with degenerate quasi-energy $E_n^l$, and a state $\phi_m^c$ with quasi-energy $E_m^c$ which is close to $E_n^l$. As  $\hbar$ is tuned, $E_m^c$ crosses $E_n^l$, but tunnel coupling makes it an avoided-crossing. In the region of this avoided crossing, one can restrict the Hilbert space, in the same spirit as in~\cite{Tomsovic/Ullmo94a}, to a subspace spanned by these three states, that we now name without label in order to simplify notation : $\phi_l$ and $\phi_r$ with energy $E$, and $\phi_c$ with energy $E'$.
In this basis one has something like
\begin{equation}
H\simeq 
\begin{pmatrix}
 E & \delta & 0 \\
 \delta & E' & \delta \\
0 & \delta & E 
\end{pmatrix}.
\end{equation}
Notice that there is no direct coupling from the left well to the right one, and only remains tunnelling from left to centre and from centre to right. In order to match with \S~\ref{subsec:husimi}, realistic values of these coefficients would be
\begin{eqnarray}
E & = & \hbar\omega_l(n+\frac{1}{2}) \label{express1}\\
E' & = & V(0)+ \hbar\omega_c(m+\frac{1}{2}) \label{express2}\\
\delta & = & \alpha \hbar e^{-A/\hbar}.
\label {express3}
\end{eqnarray}

\begin{figure}[!ht]
\center \includegraphics[width=0.8\textwidth,height=0.6\textwidth]{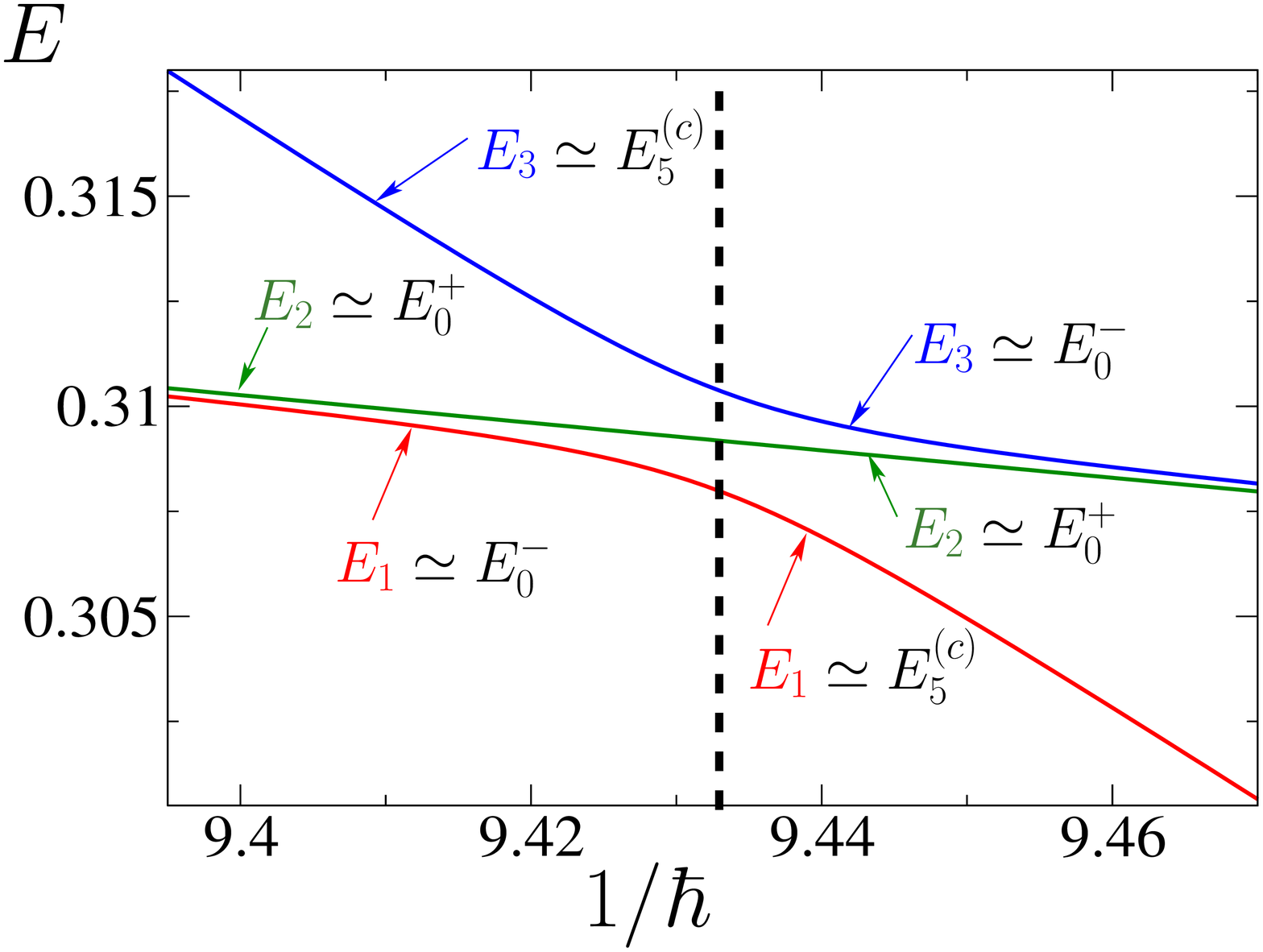}
\caption{ Avoided crossing modelised with a $3$ levels system. We have used $\omega_l=5.833$, $\omega_c=3.656$, $n=0$, $m=5$, $V(0)=-1.8225$, $\alpha=0.197$ and $A=0.34$. The vertical black dashed line indicates the value of $\hbar$ at the resonance condition $E=E'$. The numerical values are chosen according to the parameters used in figures \ref{fig:3puits_mou_pot_spectre} and \ref{fig:husimi}.}
\label{fig:avoided}
\end{figure}

It is easy to compute the corresponding eigenenergies :
\begin{eqnarray}
E_1 & = & \frac{E+E'}{2} - \frac{1}{2}\sqrt{(E-E')^2 + 8\delta^2} \\
E_2 & = & E \\
E_3 & = & \frac{E+E'}{2} + \frac{1}{2}\sqrt{(E-E')^2 + 8\delta^2}.
\end{eqnarray}
In figure \ref{fig:avoided}, we have plotted these values as a function of $1/\hbar$ by using expressions  (\ref{express1}), (\ref{express2}) and (\ref{express3}).

When $E' - E \gg \delta$, one has
\begin{eqnarray}
E_1 & \simeq & E - \frac{\delta^2}{2|E-E'|} \\
E_2 & = & E \\
E_3 & \simeq & E' + \frac{\delta^2}{2|E-E'|}.
\end{eqnarray}
This corresponds to the situation of figure \ref{fig:husimi}a.0) where only direct tunnelling is involved, and the central state has little influence. The splitting of the doublet $E_2-E_1$ is of the order $\delta^2$, that is, $e^{-2A/\hbar}$, which is basically the product of two transfer coefficients through one barrier.

When $E - E' \gg \delta$, one has
\begin{eqnarray}
E_1 &  \simeq & E' - \frac{\delta^2}{2|E-E'|} \\
E_2 & = & E \\
E_3 & \simeq & E + \frac{\delta^2}{2|E-E'|}.
\end{eqnarray}
This time, the energy of the central state $\phi_c$ has crossed the doublet, which now corresponds to $E_3-E_2$, as in the situation of figure \ref{fig:husimi}a.3).

The situation of figure \ref{fig:husimi}a.2) corresponds to the value of $\hbar$ for which the central state has resonance  $E=E'$ with the lateral ones, and one then has
\begin{eqnarray}
E_1 & = & E - \sqrt{2}\delta \\
E_2 & = & E \\
E_3 & = & E + \sqrt{2}\delta.
\end{eqnarray}
The splitting then reaches its maximum $\Delta E = 2\sqrt{2}\delta$, of the order $e^{-A/\hbar}$ which is consistent with \eqref{eq:splitting_e} and \eqref{eq:splitting_o}. Resonance with the central state thus facilitates tunnelling, by one order of magnitude, between the two non-adjacent wells.

\section{Conclusion}

While tunnelling in one dimensional time-independent systems can hardly offers some surprises, it remains
sufficiently rich to provide a warming up for dealing with much more elaborate situations that occurs in higher
dimensions (or when adding an explicit time dependency). 
In the latter cases, tunnelling frequencies or 
rates generically fluctuate by several order of magnitudes; understanding these fluctuations, that can be seen as an intricate
 overlap of resonances --- still qualitatively very different from what occurs in one-dimension---, is the aim of a
vivid field of actual researches both theoretically and experimentally \cite{Keshavamurthy/Schlagheck11a}
 and many issues still remain unclear. 
The two three-well models we have studied in this paper present the simplest bounded situation
 where tunnelling deviates from the usual purely exponential
behaviour and is indeed drastically enhanced near resonances.

\end{document}